# Spectroscopic assessment of short-term nitric acid doping of epitaxial graphene

*Ngoc Thanh Mai Tran$^{a,b}$, Swapnil M. Mhatre$^{a,c}$, Cristiane N. Santos$^{d}$, Adam J. Biacchi$^{a}$, Mathew L. Kelley$^{a,e}$, Heather M. Hill$^{a}$, Dipanjan Saha$^{a}$, Chi-Te Liang$^{c}$, Randolph E. Elmquist$^{a}$, David B. Newell$^{a}$, Benoit Hackens$^{f}$, Christina A. Hacker$^{a}$, and Albert F. Rigosi$^{a*}$*

$^a$National Institute of Standards and Technology (NIST), Gaithersburg, MD 20899, USA

$^b$Joint Quantum Institute, University of Maryland, College Park, MD 20742, USA

$^c$Graduate Institute of Applied Physics, National Taiwan University, Taipei 10617, Taiwan

$^d$Institut d'Électronique de Microélectronique et de Nanotechnologie, Université de Lille, Villeneuve-d'Ascq 59650, France

$^e$Theiss Research, La Jolla, CA 92037, USA

$^f$IMCN/NAPS Université catholique de Louvain, Louvain-la-Neuve 1348, Belgium

---

$^*$ Corresponding author. Tel: 301-975-6572. E-mail: albert.rigosi@nist.gov;

ABSTRACT

This work reports information on the transience of hole doping in epitaxial graphene devices when nitric acid is used as an adsorbent. Under vacuum conditions, desorption processes are monitored by electrical and spectroscopic means to extract the relevant timescales from the corresponding data. It is of vital importance to understand the reversible nature of hole doping because such device processing can be a suitable alternative to large-scale, metallic gating. Most measurements are performed post-exposure at room temperature, and, for some electrical transport measurements, at 1.5 K. Vacuum conditions are applied to many measurements to replicate the laboratory conditions under which devices using this doping method would be measured. The relevant timescales from transport measurements are compared with results from X-ray photoelecton spectroscopy and Fourier transform infrared spectroscopy measurements, with the latter performed at ambient conditions and accompanied by calculations of the spectra in the Reststrahlen band.

## 1. INTRODUCTION

Graphene has been the focus of many studies in recent years because of its electrical and optical properties.[1-4] Among the variety of ways it can be synthesized, epitaxially grown graphene (EG) has been shown to exhibit advantageous properties, namely its centimeter-sized growth scale and its robust quantum Hall effect.[5-9] Some applications, like resistance metrology,[10-11] require both of these advantages, and such conditions typically impose fabrication difficulties in the case where a user wishes to modulate the carrier density. In the case of EG, centimeter scale growths exhibit carrier densities that are difficult to homogenize and control, and so these kinds of samples and devices have typically undergone alternate forms of gating and doping that circumvent the need for a metallic gate that could leak through a dielectric spacer over such length scales.[12-14] The ability to apply large-scale, consistent, and predictable doping is important for any 2D-material-based device with a particular functionality, including devices with specific photovoltaic properties,[15-16] an exhibition of charge density waves,[17-18] or a potential benefit from the construction of *p-n* junctions.[19-24] Additional benefits from understanding hole doping timescales include applicability to photodetection and electron optics.[25-30]

This work details the spectroscopic assessment of timescales associated with short-term hole doping induced by exposing EG devices to nitric acid vapor. There are reports that discuss the use of nitric acid as an adsorbent,[31-33] but some elements concerning desorption timescales have yet to be fully understood. To pursue this, electrical transport data were collected at low (1.5 K) and room temperatures (298 K) to initiate a long-term temporal monitoring of the properties of the EG-based devices after being exposed to nitric acid. From these data, the desorption timescales were learned. Additionally, X-ray photoelectron spectroscopy (XPS) was employed as an additional experimental method to corroborate these observations. Lastly, Fourier transform infrared spectroscopy (FTIR) was utilized alongside additional calculations to gain insight into how changes in the Reststrahlen band can provide support for these timescales.

## 2. SAMPLE PREPARATION

*2.1 EG Growth and Device Fabrication*

EG films were grown on 4H-SiC substrates via the high-temperature sublimation method, allowing the remaining carbon atoms on the surface to restructure into a hexagonal lattice.[34] Substrate sizes of 7.6 mm × 7.6 mm were diced from 4H-SiC(0001) CREE wafers (see Notes), cleaned with a 5:1 diluted solution of hydrofluoric acid and deionized water, and coated with a diluted solution of the resist AZ 5214E to take advantage of the benefits of polymer-assisted sublimation growth (PASG).[35] The Si-face of each substrate, which has been atomically smoothed, was placed on a polished glassy carbon slab (SPI Glas 22, see Notes) to enhance large-scale single layer homogeneity. Argon gas was used to flush the graphite-lined resistive-element furnace (Materials Research Furnaces Inc., see Notes), with a final gas pressure of about 103 kPa from a 99.999 % liquid argon source. During the growth process, the furnace was held at 1900 °C for about 5 min, with heating and cooling rates of about 1.5 °C/s.

All grown EG films were vetted by inspecting their uniformity with two methods: confocal laser scanning microscopy (CLSM) and optical microscopy. Device fabrication is well documented in previous work.[36-37] To give a brief summary, steps included the deposition of a gold protection layer, photolithography for defining the Hall bar and electrical contact locations, and protection layer removal. Devices measured about 2 mm by 0.4 mm. For some devices, superconducting NbTiN was deposited as the electrical contact material as an alternate means to determine possible differences in contrast with gold contacts,[38] of which none were observed.

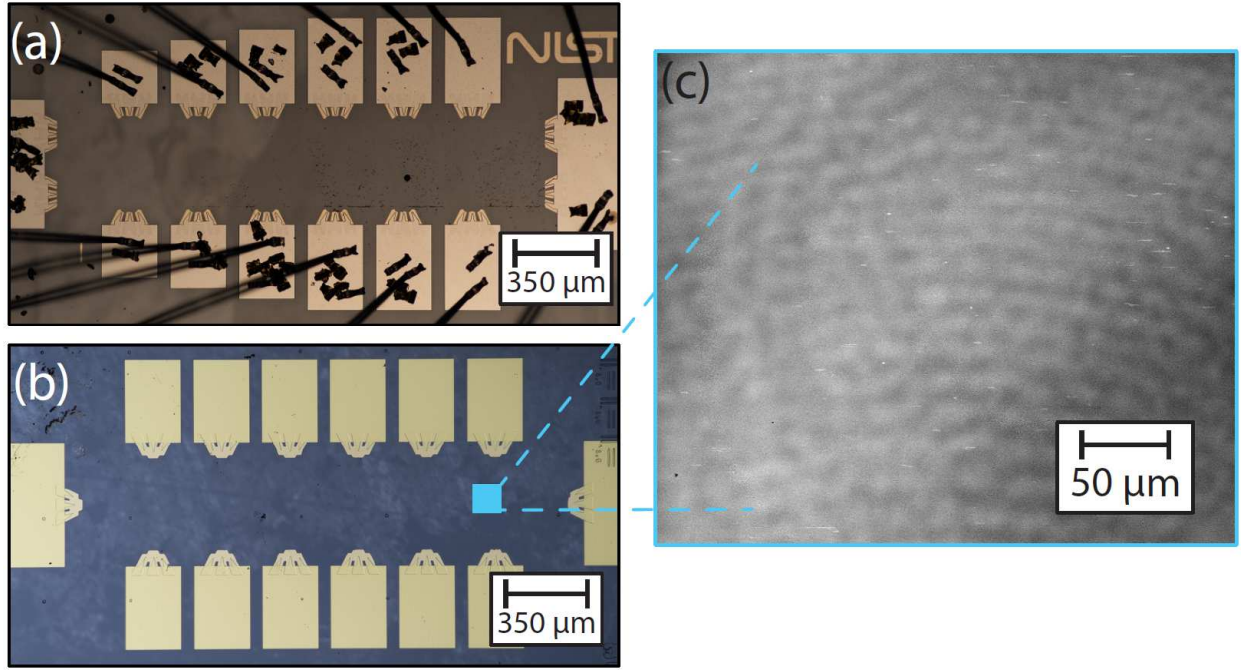

**Figure 1.** (a) An optical image of a functionalized device is shown after wire bonding the electrical contact pads. Black spots that appear within the device perimeter are clusters of oxidized chromium and do not interact with the EG layer beneath. (b) An optical image of a standard device is shown before wire bonding. The small cyan box indicates an exemplary region within which the growth quality was inspected and verified. (c) A confocal laser scanning microscopy image was collected to demonstrate the extent of homogeneous growth, with some brighter sub-micron streaks revealing minimal bilayer growth.

*2.2 EG Functionalization with Cr(CO)$_3$*

Several devices that did not serve as the control had an additional fabrication step, which was to be chemically treated via functionalization with the functional group Cr(CO)$_3$. This treatment allows the carrier density ($n_h$) to be regulated more easily without the need for a top or bottom gate. A similar process was implemented in other studies, with those on EG demonstrating a stable $n_h$ (equilibrium) close to the Dirac point.[13, 39-41] The carrier density of the functionalized EG devices following exposure to air for about one day is on the order $10^{10}$ cm$^{-2}$,[13] and this behavior of a functionalized device to asymptotically approach the Dirac point provides a valuable comparison to the control devices, especially since typical electron doping levels in EG can reach $10^{13}$ cm$^{-2}$, partly due to the buffer layer beneath the EG.[42-43]

An example set of final devices are captured in Fig. 1 (a)-(b). An optical image taken at 2.5× magnification shows a functionalized device immediately after wire bonding. The small black spots that appear scattered atop the substrate are byproduct clusters of oxidized chromium that have negligible interactions with the EG layer beneath. With Fig. 1 (b), a control device is shown before wire bonding, with the small cyan box indicating an exemplary region within which the growth quality was inspected and verified. In that same region, as shown in Fig. 1 (c), a CLSM image was acquired to further demonstrate the extent of homogeneous growth, with some brighter sub-micron streaks revealing minimal bilayer growth.

## 3. SAMPLE CHARACTERIZATION

*3.1 Quantum Electrical Transport and Nitric Acid Treatment*

For quantum Hall transport measurements and other electrical property monitoring, a Janis Cryogenics system was used (see Notes). All data were collected at temperatures of 1.5 K and magnetic fields between -9 T and 9 T. Transport data served primarily to determine the initial hole doping of each device. All samples were consistently exposed to nitric acid vapors in a standard fume hood, with each exposure taking place 3 cm from the surface of the liquid for a duration of 2 min. This duration was reported to be sufficient for $NO_3$ adsorption saturation on EG.[32] All devices were mounted onto a probe and placed under vacuum conditions (0.1 Pa).

Standard lock-in techniques were used for monitoring the longitudinal ($R_{xx}$) resistances for each device, and all source-drain currents were set to 1 µA.

*3.2 X-Ray Photoelectron Spectroscopy*

XPS measurements on control EG devices were performed using a Kratos Axis Ultra system equipped with monochromated Al Kα excitation source and a hemispherical electron analyzer collecting at an electron take-off angle of 90° with respect to the surface normal (see Notes). All XPS measurements were performed at a base pressure of 2.7 × 10$^{-7}$ Pa or less, with a pass energy of 20 eV and an energy resolution of 0.1 eV. Spectra were collected approximately 3 h after initial exposure, where the control samples had a pre-exposure $n_h$ close to the Dirac point (±10$^{10}$ cm$^{-2}$). Background spectra were collected for SiC prior to growth and EG following growth. For time-dependent measurements, a spectrum was acquired every 15 min over a period of about 12 h.

*3.3 Fourier Transform Infrared Spectroscopy and Corresponding Simulations*

Several IR spectra for all samples (that is, bare SiC, as-grown EG film, and exposed functionalized EG) were recorded with a Thermo Scientific Nicolet iS50R FTIR spectrometer (see Notes). An attenuated total reflectance module on which a sample is placed was purged with nitrogen gas prior to any measurement. The following spectrometer parameters were used: a deuterated triglycine sulfate (DTGS)-KBr detector, a range of 400 cm$^{-1}$ to 1200 cm$^{-1}$, a sample gain of 8.0, an aperture of 20, and an optical velocity of 0.475 cm/s. Data collection information for all spectra, including relevant backgrounds and reflectance are as follows: a resolution of 0.9-cm$^{-1}$, an additional 50 scans co-added for each measurement, spectral acquisition time of 43.4 s, a phase correction algorithm known as the Mertz method (in which phase errors are calculated from interferogram data and applied to the resulting complex Fourier transform), and a total time series of 150 min.

Calculations were performed to simulate the optical properties of the EG layer, specifically its complex in-plane optical conductivity including contributions from intraband and interband processes.[44] The influence of the Fermi energy ($E_F$), carrier scattering time ($\tau_\mu$), and thermal broadening ($\Gamma$) on the optical conductivity in range relevant to FTIR measurements has been analyzed in greater detail in a previous work.[44] Optimized parameters for this work include $\tau_\mu$ = 50 fs and $\Gamma$ = 18 meV. With the optical properties modeled, the reflectance of the EG layer was computed in the framework of the thin film approximation, *i.e.,* by introducing appropriate electromagnetic boundary conditions at the SiC and EG interfaces, with the total reflectance $R$ being the final output. The optical response of SiC was described by a Drude-Lorentz model.[44]

## 4. RESULTS AND DISCUSSION

*4.1 Time Constants from Carrier Density Tracking*

After doping control devices, they were placed in a cryostat and measured for their transport properties. The magnetic field sweeps at 1.5 K allowed for the determination of $n_h$. To obtain $n_h$ with transport data, one can use the relation $n_h = \frac{1}{e\left(\frac{dR_{xy}}{dB}\right)}$, where $e$ is the elementary charge, $B$ is the magnetic flux density, and $R_{xy}$ is the Hall resistance, using SI units for all quantities. The derivative of the Hall resistance is typically evaluated at low magnetic field (that is, less than 1 T). Some standard Hall measurements for a control device can be seen in Fig. 2 (a) as dashed orange and dotted green curves, whereas the same measurements for a functionalized device are shown as dashed blue and solid purple curves. The hole densities for functionalized devices are consistent after each exposure, likely because of the equilibrium near the Dirac point.[13] Control devices, unless specially prepared or pre-doped, are less likely to show narrow variation across a device as their functionalized counterparts, nor are they as likely to achieve as high an initial $n_h$. In Fig. 2 (a), the average initial $n_h$ for the example control and functionalized devices were about $3.1 \times 10^{11}$ cm$^{-2}$ and $1.1 \times 10^{12}$ cm$^{-2}$, respectively.

For time-dependent monitoring at room temperature (298 K and low pressure), $R_{xx}$ was measured for multiple adjacent longitudinal contact pairs on both control and functionalized devices. Figure 2 (b) shows examples from a single device of each type. To better understand the desorption process, a method for transforming the $R_{xx}$ data was required. To accomplish this, each device was repeatedly measured at low field and 1.5 K to obtain the unique reciprocal-like relationship between $R_{xx}$ and $n_h$. In Fig. 2 (c), for an example device, the dependence of $R_{xx}$ on $n_h$ is approximated by using a Langmuir fit to the data (with $a$, $b$, and $c$ as constants) and assumes a single-valued function:[45]

$$R_{xx} = \frac{1}{a + bn_h^{c-1}}$$

(1)

The dotted cyan perimeter is expanded in the inset, showing the region close to the Dirac point. A few points were measured to have changed polarity, indicating electron doping in the EG. Once the Langmuir fit was optimized (reduced chi-squared), the curve was used to transform $R_{xx}$ to $n_h$. Due to the consistency of functionalized devices, these transformations are shown for clarity in Fig. 2 (d). Each time-dependent $n_h$ curve was then fitted with a three-term exponential decay to accommodate for the three expected desorption processes: $NO_2$, $NO_3$, and water.[32] Three-term decays provided an optimized reduced chi-squared when compared to double- or single-term decays. The three extracted time constants ($\tau_1$, $\tau_2$, and $\tau_3$) from all collected data are 191 s ± 39 s, 7497 s ± 924 s, and $5.12 \times 10^4$ s ± $1.38 \times 10^4$ s, respectively.

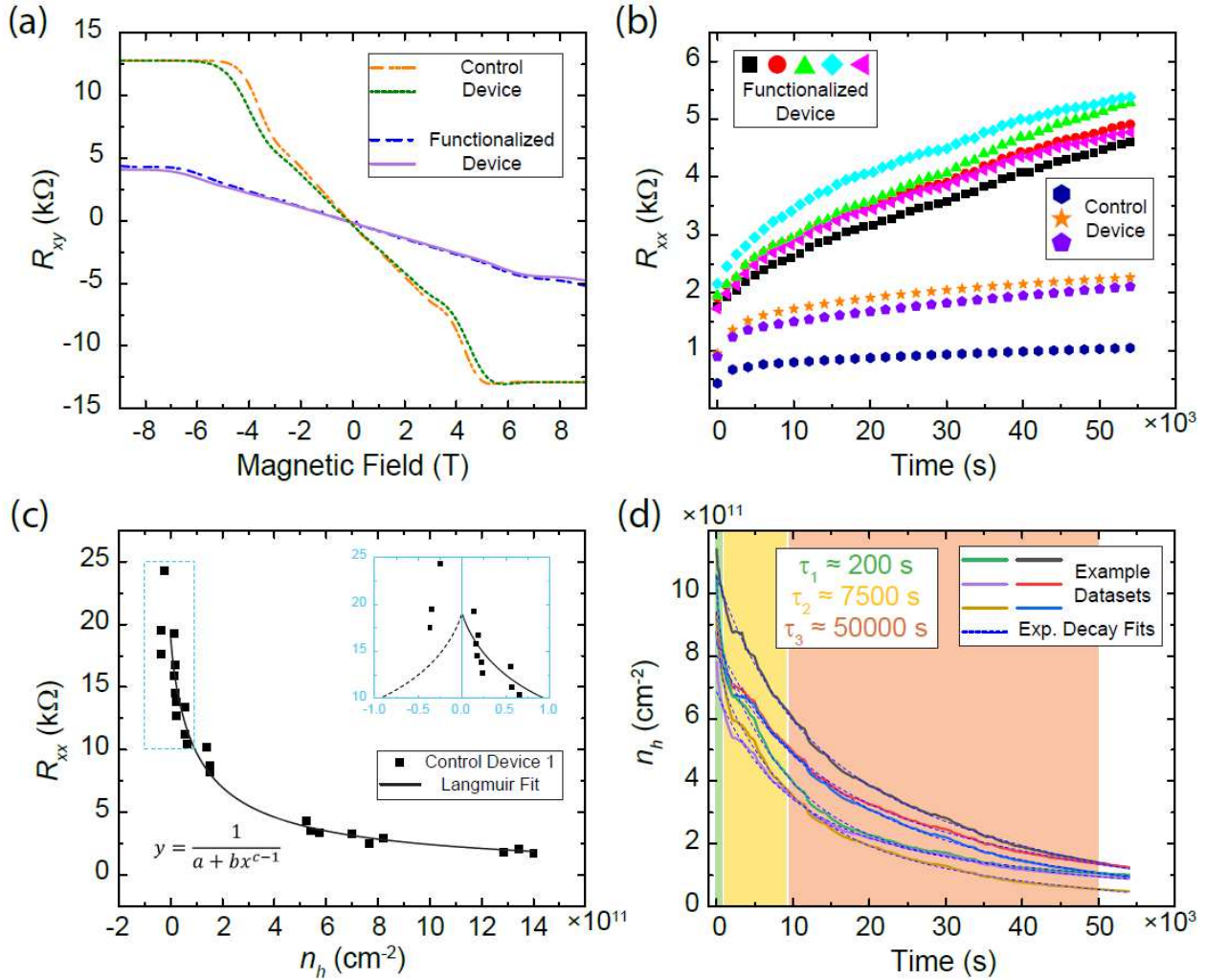

**Figure 2.** (a) Measurements of the quantum Hall effect were performed on both a control device (dashed orange and dotted green curves) and a functionalized device (dashed blue and solid purple curves). The initial hole densities ($n_h$) were calculated with the extracted slopes of the Hall curves at magnetic fields below 1 T. (b) Longitudinal resistances ($R_{xx}$) for various example longitudinal contact pairs on the control and functionalized devices were measured after each exposure to nitric acid. (c) For an example device, the dependence of $R_{xx}$ on $n_h$ is approximated by using a Langmuir fit to the data. These relationships are generally unique to each device and thus have distinct Langmuir fits. The inset magnifies the region outlined in dotted blue close to the Dirac point. A few points were measured to have changed polarity (electron doped). (d) $n_h$ is obtained from transforming $R_{xx}$, and each curve is fit with a three-term exponential decay to accommodate for three expected desorption processes. The three constants are thus extracted to be approximately 200 s, 7500 s, and 50000 s.

Some additional noteworthy details are that functionalized devices consistently started at the same initial $n_h$ after exposure (to within parts in $10^{10}$ cm$^{-2}$), providing a more predictable behavior than the control devices, whose initial $n_h$ varied by, at most, a few parts in $10^{11}$ cm$^{-2}$. It was possible for some control devices to exhibit similar levels of hole doping, but they typically required a pre-doping stage for longer exposure times. Though these samples could be checked for near-Dirac doping prior to the standardized exposure procedure, they did not perform as consistently as functionalized devices, so the latter were used to have minimal errors in the data and analysis.

To interpret the aforementioned constants, it is assumed that adsorbents have saturated the EG surface after the exposure. There are some reports discussing $NO_2$ adsorption on EG, and based on those presented data, the associated desorption times would most closely match the order of 100 s.[46-47] These observations make it reasonable to claim that $\tau_1$ comes from $NO_2$ contributions, dominant at short timescales compared to the other two terms. Since other reports indicate that the adsorption of oxygen and water on graphene occur on timescales of a few hours,[48-52] there should be few, if any, immediate competing effects. The longest timescale ($\tau_3$) may be compared with a value of similar order of magnitude in a report where water was desorbing from the EG surface.[13] According to the same report, these timescales and behaviors were reasonable given their steady state occupancies from corresponding Langmuir modeling (not related to the Langmuir fit for $R_{xx}$ data).[53] With two constants accounted for, the remaining component ($\tau_2$) may thus be attributed to the desorption of $NO_3$ in vacuum.

*4.2 Time Constant from XPS Data Analysis*

To validate the transport data, XPS data were collected, as shown in the four panels of Fig. 3 (a). The panels are arranged from top to bottom, starting with the background of an untreated EG film, first measurement ($t = 10^4$ s after exposure), $t = 1.5 \times 10^4$ s after exposure, and $t = 2.4 \times 10^4$ s after exposure. The three peak areas corresponding to $NO_3$, $NH_2$, and $NO_2$ are shaded in cyan, orange, and red, respectively and slightly offset in the vertical axis for clarity. The primary focus is on the evolution of the $NO_3$ peak, whose area was extracted for each XPS measurement and plotted as a function of time in Fig. 3 (b). Note that the horizontal time scale starts from the nitric acid exposure, and, due to standard operating procedures, several hours separate the exposure and first measurement. A single exponential decay fit was applied to obtain the time constant 5820 s ± 2319 s. The three darker points in the same panel reflect the times shown in Fig. 3 (a). As a secondary analysis, the same peak extraction was performed for $NO_2$, mainly to support the notion that dissociation is a relatively delayed phenomenon compared with $NO_2$ desorption, with a corresponding time constant of 1671 s ± 918 s.

Given that the $NO_3$ time constants from the two methods agree within their error, one can learn more about the molecular occupancy. To determine an approximate percentage of coverage of $NO_3$ that corresponds to low-pressure conditions, one may begin by assuming that both EG and $NO_3$ are homogeneous layers with uniform atomic density $\rho$, electron attenuation coefficient $\lambda$, relative sensitivity factor (RSF) $\sigma$, XPS spot area $A$, intensity $I$, and thickness $d$. The subscripts for $\lambda_{x+x}$ indicate an electron going between molecular elements. A comparable analysis is conducted in a previous work with $Cr(CO)_3$.[13] Similarly, the ratio of the intensity from the SiC substrate and EG to that of $NO_3$ may be given by Eq. 2:

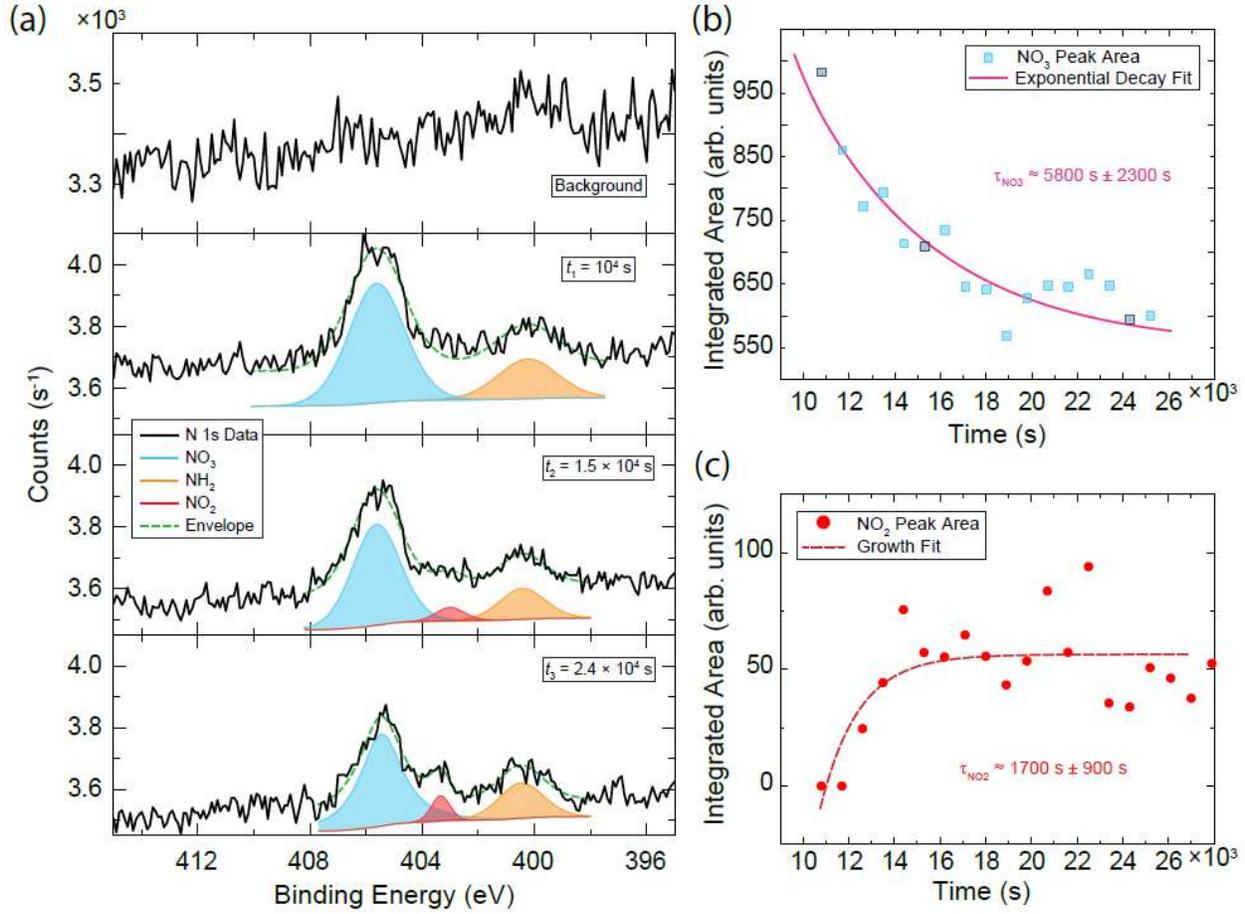

**Figure 3.** (a) XPS data are shown in four panels, as indicated from top to bottom: background (untreated EG), first measurement ($t = 10^4$ s after exposure), $t = 1.5 \times 10^4$ s after exposure, and $t = 2.4 \times 10^4$ s after exposure. The $NO_3$, $NH_2$, and $NO_2$ peak areas are shaded in cyan, orange, and red, respectively and all offset for clarity. (b) The $NO_3$ peak area was extracted for each XPS measurement and plotted as a function of time, with the time scale starting at the nitric acid exposure (several hours of achieving vacuum were necessary for the measurements). A single

exponential decay fit was applied to obtain a comparable time constant (magenta). Three darker points reflect the times shown in (a). (c) A similar analysis was performed for the $NO_2$ peak, mainly to support the notion that dissociation is a relatively delayed phenomenon compared with $NO_3$ desorption.

$$\frac{I_{Si}}{I_{NO3}} = \frac{A\sigma_{Si}\rho_{Si}\lambda_{Si+Si}\exp\left[\frac{-d_{EG}}{\lambda_{Si+EG}}\right]\exp\left[\frac{-d_{NO3}}{\lambda_{Si+NO3}}\right]}{A\sigma_{NO3}\rho_{NO3}\lambda_{NO3+NO3}\left\{1-\exp\left[\frac{-d_{NO3}}{\lambda_{NO3+NO3}}\right]\right\}}$$

(2)

This expression assumes that $\lambda_{Si+NO3} \approx \lambda_{NO3+NO3}$ and that the thickness of EG is much smaller than the attenuation distance of about 3 nm to 5 nm. Solving for $d_{NO3}$ yields:

$$d_{NO3} = \lambda_{Si+NO3}\ln\left[\frac{I_{NO3}}{I_{Si}}\frac{\sigma_{Si}\rho_{Si}\lambda_{Si+Si}}{\sigma_{NO3}\rho_{NO3}\lambda_{NO3+NO3}}+1\right]$$

(3)

The relevant parameters for Si are known from previous work.[13] The parameters for $NO_3$ required reference to additional literature, with the attenuation length being approximately 2.5 nm,[54] the RSF being approximately 0.806,[55-56] and the density $1.44 \times 10^{22}$ cm$^{-3}$. For the RSF, $\sigma_{NO3} = s\left[\frac{E_{NO3}}{E_{C1s}}\right]^{0.66}$, with the empirically derived $s$ being kept constant based on the background Si RSF. Given the counts from the data (Si as $1.5 \times 10^3$ s$^{-1}$ and $NO_3$ as 616), the estimate for the thickness of $NO_3$ is 0.05 nm (or about 14% coverage) for the first measurement, with a longer-term coverage upper bound being about 7%.

*4.3 Using Reststrahlen Band Monitoring for Ambient Conditions*

FTIR spectroscopy was utilized for monitoring transient hole behavior and was also accompanied by optical calculations in the Reststrahlen band.[44] Differential reflectance spectra were simulated using $E_F$ values between 10 meV and 80 meV, allowing a suitable metric for indicating $n_h$ to be defined, as shown in Fig. 4 (a). This metric, $\Delta_1$, quantifies the absolute value of the shift in wavenumber between the local minimum of the spectrum of interest and that of the spectrum near the Dirac point. A relationship between this metric and $E_F$ is provided in the inset, with the formula between $E_F$ and $n_h$ expressed as (with $v_F = 1.1 \times 10^8$ cm/s):

$$E_F = \hbar v_F \sqrt{\pi |n_{h,e}|} sign(n_{h,e}).$$

In Fig. 4 (b), functionalized EG was used to verify a proper calibration of the local minima between theory and experiment. Two example spectra were compared, with the red curves being at sufficiently low doping ($< 10^{10}$ cm$^{-2}$) to be effectively labeled the Dirac point. The blue curves are at an example doping of $2.2 \times 10^{11}$ cm$^{-2}$ (about 60 meV), which was induced by applying the proper annealing procedure.[13] A second metric ($\Delta_2$) was then defined using local maxima as an auxiliary measure for $n_h$. Note that this metric does not appear in the theoretical calculations but has appeared in other experimental work.[44] With the understanding of how a comparison looks between the data and theory, an inspection on a non-treated sample would be easier to analyze.

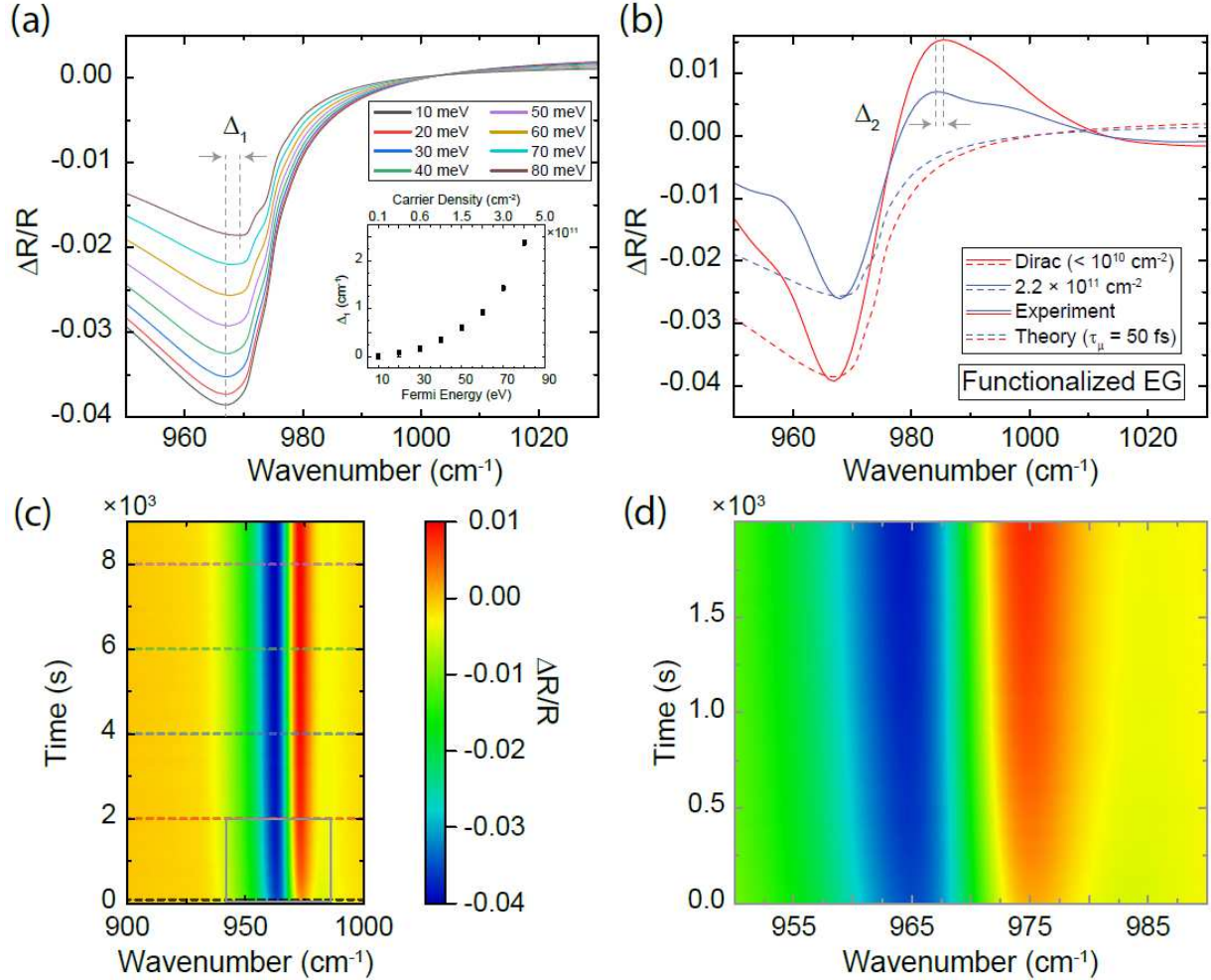

**Figure 4.** (a) The differential reflectance spectra in the range relevant for FTIR measurements are calculated in the Reststrahlen band. The Fermi energy was adjusted between 10 meV and 80 meV to find a suitable metric that could indicate $n_h$. The main metric is defined as the shift in wavenumber between the spectrum of interest and the spectrum near the Dirac point (labeled $\Delta_1$), which is shown as a function of Fermi energy in the inset. (b) Two example spectra are compared with experimental results from a functionalized sample, providing a method for calibrating $n_h$. A second metric is defined that can later be used as an auxiliary measure for $n_h$ (labeled $\Delta_2$). (c) Time-dependent reflectance contrast data acquired on a control device that was exposed to $NO_3$ are shown as a color map. The dotted lines indicate example moments later used

in the analysis. (d) The magnified region outlined by the gray box in (c) more clearly shows an evolution of the two defined metrics (showing up as blue and red, as minima and maxima, respectively).

An exemplary time-dependent map of the differential reflectance ($\Delta R/R$) was acquired for samples after their exposure to $NO_3$ and shown in Fig. 4 (c) and (d). The dotted lines indicate example moments later used in Fig. 5 (a), and the magnified region (gray box) more clearly shows the evolution of the spectra (both subfigures share the same color scale). The two metrics $\Delta_1$ and $\Delta_2$ are shown in Fig. 5 (a) for the extrema of the spectra within the Reststrahlen band at the following approximate times: 30 s (black), 2000 s (red), 4000 s (blue), 6000 s (green), and 8000 s (lavender). To create a common scale reconciling the differences in resolution between theory and experiment, a Logistic fit was simulated to establish an analytic transformation between $n_h$ and $\Delta_1$, as seen in Fig. 5 (b). The function was selected based on the optimal reduced chi-squared between the fit and the theory. Error bars in all subfigures indicate a 1σ standard deviation from either the fit calculations or experimental data.

The time-dependent data of $\Delta_1$ and $\Delta_2$ are shown in Fig. 5 (c) after exposure to $NO_3$ (blue and red, respectively). Dashed lines of identical color to the example data in Fig. 5 (a) are shown. By combining the transformation and the data from Fig. 5 (b) and (c), respectively, one then obtains a dataset for time dependent $n_h$ in Fig. 5 (d). These data are also accompanied by a single-term exponential decay, shown as a dotted orange curve. Selection of a single term was justified by the timescales established by transport in Fig. 2, suggesting that these observations are well within the regime where the desorption of $NO_3$ is the dominant contributor to the shifts in $n_h$. The shaded cyan region indicates the range of $n_h$ where the carrier polarity is highly prone to ambiguity due to Dirac point proximity. The shaded orange region indicates the extent of fitting error resulting from the experimental data error bars.

The final fitting procedure yields a $\tau_{NO3}$ of 2272 s ± 954 s. When compared to $\tau_2$ from transport, it is noted that both results differ by a factor of about 3. Two possible reasons for this difference will be explored. The first is the ambiguity of charge carrier polarity in Fig. 5 (d), which may warrant a sign change for the data at later times (> 5000 s), and such a change would modify the time constant by about 1.5 (or nearly 3500 s). Assuming an identical error bar, this modification would place the result within the error of the XPS-determined time constant. Though overlap would exist, the next reason provides more information on the discrepancy.

The second reason regards transport and XPS being performed at vacuum conditions and FTIR at ambient conditions (298 K). Regarding competing species in the framework of Langmuir adsorption,[13] the timescales on which atmospheric constituents are expected to adsorb and compete for available sites on EG are on the order of 2 h,[49-52] making further consideration of such effects necessary. Models have been established to evaluate the extent of shortened desorption lifetimes resulting from a reduction of influent concentration and competitions.[57-58] Therefore, it is not unreasonable to posit that this competition results in faster displacement of $NO_3$ as other constituents become adsorbed on the EG surface. Such displacement would artificially shorten the measured time constant obtained by the other two experimental methods. In all cases, these methods still retain their ability to provide adequate descriptions of the transient hole doping in EG.

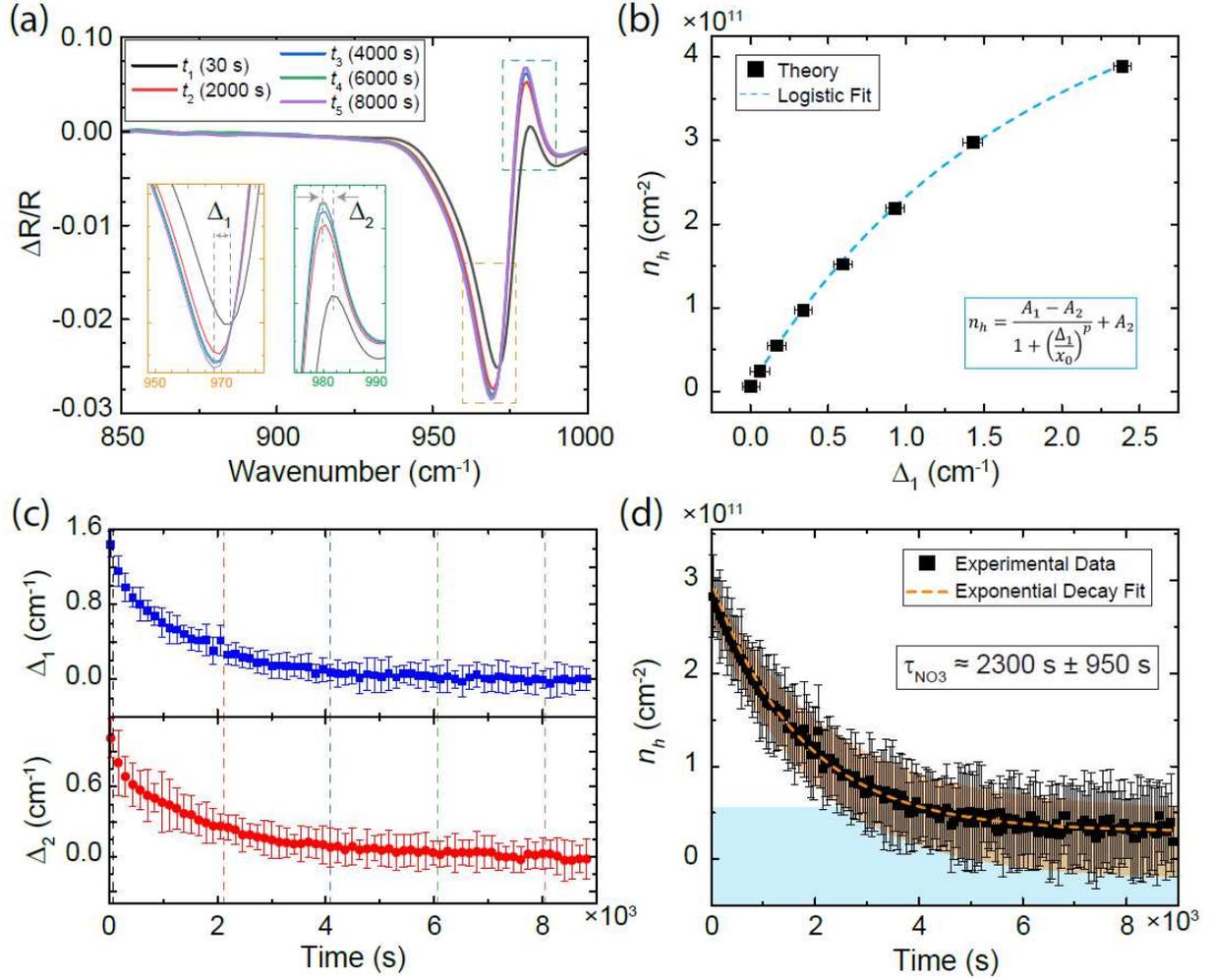

**Figure 5.** (a) Experimental FTIR differential reflectance spectra are plotted for five example moments that correspond to the dotted lines in Fig. 4 (d). The two metrics $\Delta_1$ and $\Delta_2$ are shown for the extrema of the spectra within the Reststrahlen band. (b) The simulated spectra are used to extract theoretical values of $\Delta_1$, which are then plotted to their corresponding hole densities. The relationship is fitted with a Logistic function for later transformations. (c) Time-dependent data of $\Delta_1$ and $\Delta_2$ are shown after exposure to $NO_3$. The same dashed lines appear for the example data shown in (a). (d) The transformed hole densities are plotted with time and accompanied by a single-term exponential decay (dotted orange curve). The shaded cyan region indicates the region where carrier density polarity is highly prone to ambiguity due to proximity to the Dirac

point. The shaded orange region indicates the extent of fitting error resulting from the experimental data error bars. Error bars in all subfigures indicate a 1σ standard deviation from either the fit calculations (b) or experimental data (c, d).

## 5. CONCLUSION

In this work, the transience of hole doping in epitaxial graphene devices is explored when nitric acid is used as an adsorbent. Under vacuum conditions, desorption processes are monitored by electrical and spectroscopic means to extract the relevant timescales from the corresponding data. The results suggest with vacuum conditions, a characteristic timescale for the desorption of $NO_3$ from the EG surface is about 7500 s ± 900 s according to transport, slightly lower based on XPS measurements (5800 s ± 2300 s), and substantially lower for FTIR measurements (2300 s ± 950 s), where the EG film remains at ambient temperatures (298 K) and pressures. The lattermost measurement technique was accompanied by calculations of the spectra in the Reststrahlen band to provide better understanding of the defined metrics for determining the carrier density.

**Author Contributions**

A.F.R., N.T.M.T., and S.M.M. developed the experimental design. A.F.R., N.T.M.T., S.M.M., and H.M.H. performed transport measurements. A.J.B., M.L.K, and C.A.H. performed XPS measurements and analysis. A.F.R. and C.A.H. performed FTIR measurements and analysis. D.S. and S.M.M. produced graphene samples. C.N.S. and B.N. performed calculations of the graphene FTIR response. C.-T.L., R.E.E., D.B.N., and A.F.R. provided general project support and guidance. The manuscript was written through contributions of all authors. All authors have given approval to the final version of the manuscript.

**Notes**

Commercial equipment, instruments, and materials are identified in this paper in order to specify the experimental procedure adequately. Such identification is not intended to imply recommendation or endorsement by the National Institute of Standards and Technology or the United States Government, nor is it intended to imply that the materials or equipment identified are necessarily the best available for the purpose. The authors declare no competing interests.

**Funding Sources**

Work presented herein was performed, for a subset of the authors, as part of their official duties for the United States Government. Funding is hence appropriated by the United States Congress.

ACKNOWLEDGMENT

The work of S.M.M. at NIST was made possible by arrangement with Prof. C.-T. Liang of National Taiwan University. M.K. acknowledges support from the National Institute of Standards and Technology Financial Assistance Award with Federal Award ID 70NANB19H152. B.H. (Senior Research Associate) acknowledges financial support from the F.R.S.-FNRS (Belgium). The authors thank L. S. Chao, A. L. Levy, G. J. Fitzpatrick, and E. C. Benck for their assistance with the NIST internal review process.